\documentstyle[aps,pre,amsfonts,preprint]{revtex}
\begin{document}
\draft

\title{\bf Riemannian theory of Hamiltonian chaos and Lyapunov exponents}

\author{Lapo Casetti\cite{lapo}}
\address{Scuola Normale Superiore, Piazza dei Cavalieri 7, 56126 Pisa, Italy}
\author{Cecilia Clementi\cite{cecilia}}
\address{International School for Advanced Studies (SISSA/ISAS),
via Beirut 2-4, 34014 Trieste, Italy} 
\author{Marco Pettini\cite{marco}}
\address{Osservatorio Astrofisico di Arcetri, Largo E. Fermi 5, 
50125  Firenze, Italy} 

\date {\today}

\maketitle

\begin{abstract}
A non-vanishing Lyapunov exponent $\lambda_1$ provides the very definition of 
deterministic chaos in the solutions of a dynamical system, however no 
theoretical mean of predicting its value exists.
This paper copes with the problem of analytically computing the largest 
Lyapunov exponent $\lambda_1$ for many degrees of freedom 
Hamiltonian systems as a function of $\varepsilon =E/N$, the energy per degree 
of freedom. The functional dependence $\lambda_1(\varepsilon )$  is of great 
interest because, among other reasons, it detects the existence 
of weakly and strongly chaotic regimes.
This aim - analytic computation of $\lambda_1(\varepsilon )$ - 
is successfully reached within a theoretical framework that makes use
of a geometrization of newtonian dynamics in the language of Riemannian
differential geometry. A new point of view about the origin of chaos
in these systems is obtained independently of the standard explanation based
on homoclinic intersections. Dynamical instability (chaos) is here related
to curvature fluctuations of the manifolds whose geodesics are natural
motions and is described by means of Jacobi -- Levi-Civita equation  (JLCE)  
for geodesic spread. In this paper it is shown how to derive from
JLCE an effective stability equation. Under general conditions, this effective 
equation formally describes a stochastic oscillator; 
an analytic formula for the instability growth-rate of its
solutions is worked out and applied to the Fermi-Pasta-Ulam $\beta$-model
and to a chain of coupled rotators. An excellent agreement is found between
the theoretical prediction and numeric values of $\lambda_1(\varepsilon )$
for both models.
\end{abstract}
\narrowtext

\section{Introduction}

During the last two decades or so, there has been a growing evidence of the 
independence of the two properties of {\it determinism} and {\it predictability}
of classical dynamics.
In fact, predictability for arbitrary long times requires also the 
{\it stability} of the motions with respect to variations 
-- however small -- 
of the initial conditions.

With the exception of integrable systems, the generic 
situation of classical dynamical systems describing, say, $N$ particles 
interacting through physical potentials, is {\it instability} of the 
trajectories in the Lyapunov sense. Nowadays such an instability is called
intrinsic stochasticity, or chaoticity, of the dynamics and is a consequence
of nonlinearity of the equations of motion.

Likewise any other kind of instability, dynamical instability brings about 
the exponential growth of an initial perturbation, in this case it is the 
distance between a reference trajectory and any other trajectory originating
in its close vicinity that {\it locally} grows exponentially in time.
Quantitatively, the degree of chaoticity of a dynamical system is characterized
by the largest Lyapunov exponent $\lambda_1$ that -- if positive -- 
measures the mean
instability rate of nearby trajectories averaged along a sufficiently long 
reference trajectory. The exponent $\lambda_1$ also measures the typical 
time scale of memory loss of the initial conditions.

Let us remember that if
\begin{equation}
\dot x^i = X^i(x^1\dots x^N)
\label{eq1}
\end{equation}
is a given dynamical system, i.e. a realisation in local coordinates of a
one-parameter group of diffeomorphisms of a manifold $M$, that is of
$\phi^t: M\rightarrow M$, and if we denote by
\begin{equation}
\dot\xi^i = {\cal J}^i_k[ x(t)]\, \xi^k
\label{eq2}
\end{equation}
the usual tangent dynamics equation, i.e. the realisation of the mapping
$d\phi^t: TM\rightarrow TM$, where $TM$ is the tangent bundle of $M$ and
$[{\cal J}^i_k]$ is the Jacobian matrix of $[X^i]$, then the largest Lyapunov
exponent $\lambda_1$ is defined by
\begin{equation}
\lambda_1 = {\displaystyle\lim_{t\rightarrow\infty}}~\frac{1}{t}\ln \frac
{\Vert\xi (t)\Vert}{\Vert\xi (0)\Vert}
\label{eq3}
\end{equation}
and, by setting $\Lambda [x(t),\xi (t)]= \xi^T\,{\cal J}[x(t)]\, 
\xi /\,\xi^T\xi\equiv\xi^T {\dot\xi} /\xi^T \xi 
=\frac{1}{2}\frac{d}{dt}\ln (\xi^T\xi )$, 
this can be formally expressed as a time average
\begin{equation}
\lambda_1 ={\displaystyle\lim_{t\rightarrow\infty}}~\frac{1}{2t}\int_0^t 
\,d\tau \,\Lambda [x(\tau ), \xi (\tau )]~~~.
\label{eq4}
\end{equation}
Even though $\lambda_1$ is the most important indicator of chaos of classical 
\cite{quantlyap} dynamical systems, it is used only as a diagnostic tool in 
numerical simulations. With the exception of a few simple 
discrete-time systems (maps of the interval), no theoretical method exists
to compute $\lambda_1$ \cite{Gozzi}. 
This situation reveals that a satisfactory theory
of deterministic chaos is still lacking, at least for systems of physical
relevance.

In the conventional theory of chaos, dynamical instability is caused by
homoclinic intersections of perturbed separatrices, however this theory 
has many problems: {\it i)}
it needs action-angle coordinates, {\it ii)} it works in conditions of
weak perturbation of an integrable system, {\it iii)} to compute quantities
like Mel'nikov integrals one needs the analytic expressions of the unperturbed 
separatrices: at large $N$ this is hopeless, moreover the generalization
of Poincar\'e-Birkhoff theorem is still problematic at $N>2$; {\it iv)}
finally, there is no computational relationship 
between homoclinic intersections
and Lyapunov exponents. Therefore this theory seems not adequate to treat chaos
in Hamiltonian systems with many degrees of freedom at arbitrary degree of
nonlinearity, with potentials that can be hardly transformed in action-angle
coordinates, not to speak of accounting for phenomena like the transition
from weak to strong chaos in Hamiltonian systems 
\cite{{PettiniLandolfi},{PettiniCerruti}}.
Motivated by the need of understanding this transition from weak to strong
chaos, we have recently proposed 
\cite{Pettini,CasettiPettini,Lyap,CerrutiPettini,CerrutiPettini1,ValdetPettini}
to tackle
Hamiltonian chaos in a theoretical framework different from that of homoclinic 
intersections. This new method makes use
of the well-known possibility of formulating Hamiltonian dynamics in the 
language of Riemannian geometry so that the stability or instability of a 
geodesic
flow depends on curvature properties of some suitably defined manifold.

In the early 1940s, N. S. Krylov already got a hold of the potential interest 
of this differential-geometric framework to account for dynamical instability
and hence for phase space mixing \cite{Krylov}. The follow-up of his intuition
can be found in abstract ergodic theory \cite{Sinai} and in a very few 
mathematical works concerning the ergodicity of geodesic flows of physical 
interest \cite{Knauf,Gutzwiller}. However, Krylov's work 
did not entail anything useful
for a more general understanding of chaos in nonlinear newtonian dynamics
because one soon hits against unsurmountable mathematical obstacles.
By filling certain mathematical gaps with numerical investigations, 
these obstacles have been overcome and a rich scenario emerged about the
relationship between stability and curvature 

Based on the so-obtained information, the present paper aims at bringing a
substantial contribution to the development of a Riemannian theory 
of Hamiltonian chaos. The new contribution consists of a method to 
analytically compute 
the largest Lyapunov exponent $\lambda_1$ for physically meaningful Hamiltonian
systems of arbitrary large number of degrees of freedom. A preliminary 
and limited account of the results presented here can be found in 
Ref. \cite{Lyap}.

The paper is organized as follows: Section \ref{sezione2} is a sketchy  
presentation of the geometrization of  newtonian dynamics; Section 
\ref{sezione3} contains
the derivation of an effective stability equation from Jacobi -- Levi-Civita 
equation for geodesic spread and an analytic formula for $\lambda_1$;
Section \ref{sezione4} contains the application of the general result to the
practical computation of $\lambda_1$ in the Fermi-Pasta-Ulam $\beta$-model
and in a chain of coupled rotators. Some concluding remarks are presented in
Section \ref{sezione5}.

\section{Geometrization of newtonian dynamics}
\label{sezione2}
Let us briefly recall 
how newtonian dynamics can be rephrased in the language of
Riemannian geometry. We shall deal with standard autonomous systems, i.e.
described by the Lagrangian function
\begin{equation}
{\cal L} = T - V = 
\frac{1}{2} a_{ij} \dot q_i \dot q_j - V(q_1,\ldots,q_N)~, 
\label{eq5}
\end{equation}
so that the Hamiltonian function ${\cal H} = T + V\equiv E$ is a constant of motion.

According to the principle of stationary action -- in the form of Maupertuis --
among all the possible isoenergetic paths $\gamma (t)$ with fixed end points,
the paths that make vanish the first variation of the action functional
\begin{equation}
{\cal A} = \int_{\gamma(t)} p_i \, dq_i 
         = \int_{\gamma(t)} \frac{\partial {\cal L}}{\partial \dot q_i}
	  \, \dot q_i\, dt 
\label{action_M}
\end{equation}
are natural motions.

As the kinetic energy $T$ is a homogeneous function of degree two, we have
$2T = \dot q_i \partial {\cal L}/{\partial \dot q_i}~,$ and Maupertuis' 
principle reads
\begin{equation}
\delta{\cal A} = \delta \int_{\gamma(t)} 2T \, dt = 0~.
\label{M_principle}
\end{equation}
The configuration space $M$ of a system with $N$ degrees of freedom is an
$N$-dimensional differentiable manifold and the lagrangian coordinates
$(q_1,\ldots,q_N)$ can be used as local coordinates on $M$. The manifold $M$ 
is naturally given a proper Riemannian structure. In fact, let us consider 
the matrix
\begin{equation}
g_{ij} = 2[E - V(q)] a_{ij}
\end{equation}
so that (\ref{M_principle}) becomes
\begin{equation}
\delta \int_{\gamma(t)} 2T \, dt = 
\delta \int_{\gamma(t)} \left( g_{ij} \dot q^i \dot q^j \right)^{1/2} \, dt =
\delta \int_{\gamma(s)} ds\, = 0 ~,   
\end{equation}
thus natural motions are geodesics of $M$, provided we define
$ds$ as its arclength. The metric tensor $g_J$ of $M$ is then defined by
\begin{equation}
g_J = g_{ij} \, dq^i \otimes dq^j
\end{equation}
where $(dq^1,\ldots,dq^N)$ is a natural base of $T^*_q M$ - the cotangent space
at the point $q$ - in the local chart
$(q^1,\ldots,q^N)$. This is known as Jacobi (or kinetic energy) metric.
Denoting by $\nabla$ the canonical Levi-Civita connection, the geodesic 
equation
\begin{equation}
\nabla_{\dot\gamma} \dot\gamma = 0
\end{equation}
becomes, in the local chart $(q^1,\ldots,q^N)$,
\begin{equation}
\frac{d^2 q^i}{ds^2} + \Gamma^i_{jk} \frac{dq^j}{ds} \frac{dq^k}{ds} = 0~,
\label{eq_geodesics_loc}
\end{equation}
where the Christoffel coefficients are the components of $\nabla$ defined by
\begin{equation}
\Gamma^i_{jk} = \langle dq^i, \nabla_j e_k \rangle = \frac{1}{2} g^{im} 
\left( \partial_j g_{km} + \partial_k g_{mj} - \partial_m g_{jk} \right)~,
\label{Gamma}
\end{equation}
where $\partial_i = \partial/\partial q^i$. Without loss of generality
consider $g_{ij} = 2[E-V(q)]\delta_{ij}$, from Eq. (\ref{eq_geodesics_loc}) 
we get 
\begin{equation}
\frac{d^2 q^i}{ds^2} + \frac{1}{2(E - V)}
\left[2 \frac{\partial (E-V)}{\partial q_j} \frac{dq^j}{ds} \frac{dq^i}{ds} 
- g^{ij} \frac{\partial (E-V)}{\partial q_j} 
g_{km}\frac{dq^k}{ds} \frac{dq^m}{ds} \right] = 0~,
\end{equation}
and, using $ds^2 = 2(E-V)^2\, dt^2$, we can easily verify that these equations
yield
\begin{equation}
\frac{d^2 q^i}{dt^2} = - \frac{\partial V}{\partial q_i}~~~~i=1,\dots ,N~.
\end{equation} 
which are Newton equations.

As already discussed elsewhere \cite{{Pettini},{CasettiPettini}}, 
there are other 
possibilities to associate a Riemannian manifold to a standard Hamiltonian
system. Among the others we mention a structure, defined by Eisenhart 
\cite{Eisenhart}, that will be used in the following for computational reasons.
In this case the ambient space is an enlarged configuration space-time
$M \times {{\Bbb R}}^2$, with local coordinates
$(q^0, q^1,\ldots,q^N,q^{N+1})$,  with $(q^1,\ldots,q^N)\in M$, 
$q^0\in{\Bbb R}$
is the time coordinate, 
$q^{N+1}\in{\Bbb R}$ is a coordinate closely related to 
the action; Eisenhart defines a pseudo-Riemannian non-degenerate metric 
$g_{{}_E}$ on $M\times{\Bbb R}^2$ as
\begin{equation}
ds_{{}_E}^2= g_{\mu\nu}\, dq^{\mu} \otimes dq^{\nu} = 
a_{ij} \, dq^i \otimes dq^j -2V(q)\, dq^0 \otimes dq^0 
+ dq^0 \otimes dq^{N+1} + dq^{N+1} \otimes dq^0~.
\label{g_E}
\end{equation}
Natural motions are now given by the canonical projection $\pi$ of the
geodesics of $(M\times{\Bbb R}^2, g_E)$ on configuration space-time:
$\pi : M\times{\Bbb R}^2\rightarrow M\times{\Bbb R}$. However, among all the 
geodesics of $g_E$ we must consider only those for which the arclength is
positive definite and given by
\begin{equation}
ds^2 = g_{\mu\nu} dq^\mu dq^\nu = 2C^2 dt^2~,
\label{par_affine}
\end{equation}
or, equivalently, we have to consider only those geodesics such that the 
coordinate $q^{N+1}$ evolves according to
\begin{equation}
q^{N+1} = C^2 t + C^2_1 - \int_0^t {\cal L}\, d\tau~,
\label{qN+1}
\end{equation}
where $C$ and $C_1$ are real constants. Since the values of these constants
are arbitrary, we fix $C^2 = 1/2$ in order that $ds^2 = dt^2$ along a
physical geodesic.
For a diagonal kinetic energy matrix $a_{ij} = \delta_{ij}$, the non vanishing 
components of the connection $\nabla$ are simply
\begin{equation}
\Gamma^i_{00} = - \Gamma^{N+1}_{0i} = \partial_i V~, 
\end{equation}
therefore it is easy to check that also the geodesics of $g_{{}_E}$ yield Newton
equations together with the differential versions of Eq. (\ref{qN+1}) and of
$q^0=t$ (details can be found in \cite{{Pettini},{CasettiPettini}}).

\section{Geometric description of dynamical instability}
\label{sezione3}
The actual interest of the Riemannian formulation of dynamics stems from the
possibility of studying the instability of natural motions through the
instability of geodesics of a suitable manifold, a circumstance that has 
several advantages. First of all a powerful mathematical tool exists to
investigate the stability or instability of a geodesic flow: the Jacobi --
Levi-Civita equation (JLC) for geodesic spread. The JLC equation describes
covariantly how nearby geodesics locally scatter and it is a familiar object
both in Riemannian geometry and theoretical physics (it is of fundamental
interest in experimental General Relativity). Moreover the JLC equation 
relates the stability or instability of a geodesic flow with {\it curvature}
properties of the ambient manifold, thus opening a wide and largely 
unexplored field
of investigation of the connections among geometry, topology and geodesic
instability, hence chaos.
\smallskip
\subsection{Jacobi - Levi Civita equation for geodesic spread}

A {\em congruence of geodesics} is defined as a family of geodesics
$\{ \gamma_\tau(s) = \gamma(s,\tau) \, | \, \tau \in {{\Bbb R}} \}$ that, 
originating in some neighbourhood ${\cal I}$ of any given point of 
a manifold, are differentiably
parametrized by some parameter $\tau$. Choose a reference
geodesic $\bar\gamma (s,\tau_0)$, denote by $\dot\gamma(s)$ the field of
vectors tangent at $s$ to $\bar\gamma$ and denote by $J(s)$ the field of
vectors tangent at $\tau_0$ to the curves $\gamma_s(\tau)$ at fixed $s$.
The field $J=(\partial \gamma/\partial\tau )_{\tau_0}$ 
is known as  {\em geodetic separation field} and it has the property:
${\cal L}_{\dot\gamma}J=0$, where ${\cal L}$ is the Lie derivative. 
Locally we can
measure the distance between two nearby geodesics by means of $J$. 

The evolution of the geodetic separation field $J$ conveys information
about stability or instability of the reference geodesic $\bar\gamma$,
in fact, if $\Vert J\Vert$ 
exponentially grows with $s$ then the geodesic is unstable 
in the sense of Lyapunov, otherwise it is stable.

The evolution of $J$ is described by \cite{DoCarmo}
\begin{equation}
\frac{\nabla^2 J(s)}{ds^2} + R (\dot{\gamma}(s),J(s))\,\dot{\gamma}(s) = 0~,
\label{eq_jacobi}
\end{equation}
known as Jacobi -- Levi-Civita (JLC) equation. Here $J(s)\in T_{\gamma (s)}M$;
$R(X,Y)=\nabla_X\nabla_Y - \nabla_Y\nabla_X - \nabla_{[X,Y]}$ is the 
Riemann-Christoffel curvature tensor; $\dot\gamma =d\gamma /ds$; 
$\nabla /ds$ is the covariant derivative and $\gamma (s)$ is a normal
geodesic, i.e. such that $s$ is the length. In the following we assume that
$J(s)$ is normal, i.e. $\langle J, \dot\gamma\rangle =0$.
This equation relates the stability or instability of nearby geodesics to
the curvature properties of the ambient manifold. If the ambient manifold
is endowed with a metric (e.g. Jacobi or Eisenhart) derived from the Lagrangian
of a physical system,  
then stable or unstable (chaotic) motions will depend on the curvature
properties of the manifold. Therefore it is reasonable to guess that some
{\it average} global geometric property will provide information, at least,  
about an {\it average} degree of chaoticity of the dynamics independently of 
the knowledge of the trajectories, that is independently of the numerical 
integration of the equations of motion.

In local coordinates the JLC equation (\ref{eq_jacobi}) reads as
\begin{equation}
\frac{\nabla^2 J^i}{ds^2} + 
R^i_{~jkl} \frac{dq^j}{ds}{J^k}\frac{dq^l}{ds} = 0~,
\label{eq_jacobi_geo}
\end{equation}
where $R^i_{~jkl} = \langle dq^i, R(e_{(k)},e_{(l)}) e_{(j)} \rangle$ 
are the components of the curvature tensor, and the covariant derivative is 
$(\nabla J^i/ds)= dJ^i/ds + \Gamma^i_{jk}J^k dq^j/ds$.
There are ${\cal O}(N^4)$ of such components, $N = \dim M$, 
therefore -- even if this number
can be considerably reduced by symmetry considerations -- 
equation (\ref{eq_jacobi_geo}) appears untractable already at rather
small $N$. It is worth mentioning that some exception exists. Such is the case
of {\it isotropic} manifolds for which (\ref{eq_jacobi_geo}) can be reduced to
the simple form
\begin{equation}
\frac{\nabla^2 J^i}{ds^2} + K J^i = 0~~~i=1,\dots ,N~,
\label{JLC_isotrop}
\end{equation}
where $K$ is the constant value assumed throughout the manifold by the 
sectional curvature.

The sectional curvature of a manifold is the $N$-dimensional generalization 
of the gaussian curvature of two-dimensional surfaces of ${\Bbb R}^3$. Consider
two arbitrary vectors $X,Y \in T_x M$, where $x\in M$ is an arbitrary point
of $M$, and define
\begin{equation}
\Vert X \wedge Y\Vert = (\Vert X\Vert^2 \Vert Y\Vert^2 - \langle X, Y\rangle )
^{1/2}
\label{X^Y}
\end{equation}
if $\Vert X\wedge Y\Vert \neq 0$ the vectors $X,Y$ span a two-dimensional
plane $\pi\subset T_x M$, then the sectional curvature at $x$ relative to 
the plane $\pi$ is defined by
\begin{equation}
K(X, Y) = K(x, \pi )= \frac{\langle R(Y, X) X, Y\rangle}{\Vert
X\wedge Y\Vert^2}
\label{kappa2}
\end{equation}
which is only a property of $M$ at $x$ independently of $X,Y\in \pi$
(Gauss' theorema egregium).
For an isotropic manifold $K(x, \pi )$ is also independent of the choice
of $\pi$ and thus, according to Schur's theorem, $K$ turns out also
independent of $x\in M$.

Unstable solutions of the equation (\ref{JLC_isotrop}) are of the form
\begin{equation}
J(s)= w(0)(-K)^{-1/2} \sinh \left(\sqrt{ - K}\, s\right)~~,
\label{cosh}
\end{equation}
once the initial 
conditions are assigned as $J(0) = 0$ and $dJ(0)/ds = w(0)$ and $K < 0$.
In abstract ergodic theory geodesic flows on compact manifolds of constant
negative curvature have been considered in classical works
\cite{Anosov}.
In this case the quantity $\sqrt{- K}$ -- uniform on the manifold -- 
measures the
degree of instability of nearby geodesics.

While Eq.  (\ref{JLC_isotrop}) holds true only for constant curvature 
manifolds, a similar form of general validity can be obtained for JLC equation
at $N=2$.

In this low-dimensional case Eq. (\ref{eq_jacobi_geo}) is exactly rewritten
as
\begin{equation}
\frac{d^2 J}{ds^2} + \frac{1}{2} {\cal R}(s) J =0 ~,
\label{JLCN=2}
\end{equation}
where a parallely transported frame is used and
${\cal R}(s)$ is the scalar curvature. Using Jacobi metric one finds
($N=2$): ${\cal R}=\triangle V/W^2 + (\nabla V)^2/W^3$, with $W=E-V$, so that
for smooth and binding potentials ${\cal R}$ can be negative only where
$\triangle V < 0$, i.e nowhere for nonlinearly coupled oscillators as
described by the H\'enon-Heiles model \cite{CerrutiPettini1} or for quartic
oscillators \cite{ValdetPettini}. $\triangle V <0$ is only possible if the
potential $V$ has inflection points.

Recent detailed analyses of two-degrees of freedom systems
\cite{{CerrutiPettini1},{ValdetPettini}} have shown that  chaos 
can be produced
by {\it parametric instability}
due to a fluctuating positive curvature along the geodesics.

 Let us remember that parametric instability is a generic 
property of dynamical systems with parameters that are periodically or 
quasi-periodically varying in time, even if for each value of the varying
parameter the system has stable solutions \cite{nayfeh}. 
A harmonic oscillator with
periodically modulated frequency, described by the Mathieu equation, is
perhaps the prototype of such a parametric instability mechanism.

Numerical simulations 
have shown that all the informations about order and chaos
obtained by standard means (Lyapunov exponent and Poincar\'e sections) are
fully retrieved by using Eq. (\ref{JLCN=2}). As in the case of tangent dynamics,
Eq.  (\ref{JLCN=2}) has to be computed along a reference geodesic (trajectory).

Let us now cope with the large $N$ case. 
It is convenient to rewrite the JLC equation (\ref{eq_jacobi_geo})
in the following form 
\begin{equation}
\frac{\nabla^2 J(s)}{ds^2} + 
\frac{1}{N-1}\left[ \text{Ric} (\dot\gamma(s),\dot\gamma(s))\,
J(s)\,-\, \text{Ric} (\dot\gamma(s),J(s))\,\dot\gamma(s)
\right] \, +\, W(\dot\gamma(s), J(s))\,\dot\gamma(s) = 0~,
\label{eq_jacobi_weyl}
\end{equation}
where $W$ is the Weyl projective curvature tensor whose components $W^i_{jkl}$ 
are given by \cite{goldberg}
\begin{equation}
W^i_{~jkl} = R^i_{~jkl} - \frac{1}{N-1} (R_{jl} \delta^i_k - R_{jk} 
\delta^i_l)~,
\label{weyl}
\end{equation}
and Ric is the Ricci curvature tensor of components $R_{ij} = R^m_{~imj}$.
Weyl's projective tensor $W$ (not to be confused with Weyl's {\it conformal} 
curvature tensor) measures the deviation from isotropy of a given manifold.
For an isotropic manifold $W^i_{jkl}=0$, and we recognize in 
(\ref{eq_jacobi_weyl}) equation
(\ref{JLC_isotrop}), in fact in this case $R_{jl}\dot q^j\dot q^l/(N-1)$ is 
just the constant value of sectional curvature. Remind that the Ricci curvature 
at $x\in M$ is
$K_R(X_{(b)}) = R_{jl}X_{(b)}^iX_{(b)}^l = \sum_{a=1}^{N-1}K(X_{(b)},
X_{(a)})$ where $X_{(1)},\dots ,X_{(N)}$ form an orthonormal basis of $T_xM$.
Hence we understand that Eq. (\ref{eq_jacobi_weyl}) retains the structure of 
Eq. (\ref{JLC_isotrop}) up to its second term that now has the meaning of a
mean sectional curvature averaged, at any given point, over the independent 
orientations of the planes spanned by $X_{(a)}$ and $X_{(b)}$; this mean 
sectional curvature is no longer constant along $\gamma(s)$. The last term
of (\ref{eq_jacobi_weyl}) accounts for the local degree of anisotropy of the 
ambient manifold.

Let us now consider the following decomposition for the Jacobi field $J$
\begin{equation}
J(s)=\sum_i J_i(s)\,e_{(i)}(s)
\end{equation}
where $\{ e_{(1)}\dots e_{(N)}\}$ is an orthonormal system of
parallely transported vectors. In this reference frame it is
\begin{equation}
{{\nabla^2J}\over{ds^2}} = \sum_i{{d^2 J_i}\over{ds^2}}\,e_{(i)}(s)
\end{equation}
and the last term of (\ref{eq_jacobi_weyl}) is
\begin{eqnarray}
W(\dot\gamma, J)\dot\gamma&=&\sum_j \langle W(\dot\gamma,J)\dot\gamma ,
e_{(j)}\rangle\, e_{(j)} \nonumber \\
 &=&\sum_j\langle W(\dot\gamma, \sum_i J_i e_{(i)})\dot\gamma, e_{(j)}
\rangle\, e_{(j)}\\
 &=&\sum_{ij}\langle W(\dot\gamma , e_{(i)})\dot\gamma, e_{(j)}
\rangle\, J_i\, e_{(j)}~~, \nonumber
\end{eqnarray}
the same decomposition applies to the third term of Eq. (\ref{eq_jacobi_weyl})
which is finally rewritten as
\begin{equation}
{{d^2 J_j}\over{ds^2}}+ k_R(s)\, J_j + \sum_i\, (w_{ij} + r_{ij})\, J_i = 0
\label{sistemaHill}
\end{equation}
where $k_R = K_R/(N-1)$, 
$w_{ij}=\langle W(\dot\gamma , e_{(i)})\dot\gamma , e_{(j)}\rangle$ and
$r_{ij}=\langle\text{Ric}(\dot\gamma , e_{(i)})\dot\gamma , e_{(j)}
\rangle /(N-1)$.
Of course $k_R$ is independent of the coordinate system.
The elements $w_{ij}$ still depend on the dynamics and on the behavior of the
vectors $e_{(k)}(s)$, thus, in order to obtain a stability equation, for the
geodesic flow, that depends only on average curvature properties of the
ambient manifold, we try to conveniently  approximate the $w_{ij}$.
To this purpose define at any point $x\in M$ the trilinear mapping
$R^\prime:T_xM\times T_xM\times T_xM \rightarrow T_xM$ by
\begin{equation}
\langle R^\prime (X,Y,U),Z\rangle = \langle X,U\rangle\langle Y,Z\rangle -
\langle Y,U\rangle\langle X,Z\rangle
\label{trilinear}
\end{equation}
for all $X,Y,U,Z\in T_xM$. It is well known \cite{DoCarmo} that, if and only 
if $M$ is isotropic then $R=K_0R^\prime$, where $R$ is the Riemann curvature 
tensor of $M$ and $K_0$ is the constant sectional curvature.

Let us now assume that the ambient manifold is {\it quasi-isotropic}, i.e.
that it looks like an isotropic manifold after a coarse-graining that smears 
out all the  metric fluctuations, and let us formulate this 
assumption by putting $R\approx K(s) R^\prime$ and  $\text{Ric}\approx K(s) g$,
although $K(s)$ is no longer
a constant. Now we use (\ref{trilinear}) to find $w_{ij}\approx\delta K(s)
[\langle\dot\gamma ,\dot\gamma\rangle\langle e_{(i)},e_{(j)}\rangle -
\langle e_{(i)},\dot\gamma\rangle\langle \dot\gamma ,e_{(j)}\rangle ]$, then 
we use $\text{Ric}\propto g$ and $g(\dot\gamma ,J) = 0$ to find $r_{ij}=0$ thus 
Eq. (\ref{sistemaHill}) becomes
\begin{equation}
{{d^2 J_j}\over{ds^2}}+ k_R(s)\, J_j + \delta K(s)\,J_j = 0~,
\label{eq_Hill}
\end{equation}
by $\delta K(s) = K(s) - {\overline K}$ we denote the local deviation of
sectional curvature from its coarse-grained value ${\overline K}$, thus
$\delta K(s)$ measures the fluctuation of sectional curvature along a geodesic
due to the local deviation from isotropy. 
The problem is that $\delta K(s)$ still depends
on a moving plane $\pi (s)$ determined by $\dot\gamma (s)$ and $J(s)$. In 
order to get rid of this dependence, consider that if $x\in M$ is an 
isotropic point then the components of the Ricci tensor are 
$R_{lh}=(N-1)K(x)g_{lh}$ and the scalar curvature is
$R=N(N-1)K(x)$; with these quantities one constructs the Einstein 
tensor $G_{lh}=R_{lh}-\frac{1}{2}g_{lh}R$ whose divergence vanishes 
identically ($G_{lh\vert l}=0$) so that it is immediately found that, if 
a manifold consists entirely of isotropic points, 
then $\partial K(x)/\partial x^l=0$ 
and so $\partial K_R(x)/\partial x^l=0$, 
i.e. the manifold is a space of constant curvature (Schur's theorem 
\cite{DoCarmo}).
Conversely, the local variation of Ricci curvature detects the
local loss of isotropy, thus a reasonable approximation of the {\it average} 
variation $\delta K(s)$ along a geodesic may be given by the 
variation of Ricci curvature. 

Next let us model $\delta K(s)$ along a geodesic by a stochastic process. 
In fact $K(s)$ is obtained by summing a large number of 
terms, each one depending on different combinations of the components of
$J$ and on the the coordinates $q^i$, moreover, unless we tackle an integrable
model, the dynamics is always chaotic and the functions $q^i(s)$ behave 
irregularly. By invoking a central-limit-theorem argument, at large $N$, 
$\delta K(s)$ is expected to behave, in first approximation, as a gaussian
stochastic process.
More generally, the probability distribution ${\cal P}(\delta K)$ may be
other than gaussian and in practice it could be 
determined by computing its cumulants along a geodesic $\gamma (s)$.

Now we make quantitative the previous statement -- about using the variation of
Ricci curvature along a geodesic to estimate $\delta K(s)$ -- by putting
\begin{equation}
{\cal P}(\delta K)\simeq{\cal P}(\delta K_R)~.
\label{probabilita}
\end{equation}
Both $\delta K$ and $\delta K_R$ are zero mean variations, so the first 
moments vanish; according to (\ref{probabilita}) the following relation 
for the second moments will hold
\begin{equation} 
\langle [K(s) - {\overline K}]^2\rangle_s \simeq
\frac{1}{N-1} \langle [K_R(s) - \langle K_R \rangle_s]^2 \rangle_s ~,
\label{momento2}
\end{equation}
where $\langle \cdot \rangle_s$ stands for proper-time average along a 
geodesic $\gamma(s)$. 
Let us comment about the numerical factor in the r.h.s. of (\ref{momento2})
where a factor $\frac{1}{N^2}$ might be expected. At increasing $N$ the
mean square fluctuations of $k_R$ drop to zero as $\frac{1}{N}$ because
$k_R$ is the mean of independent quantities, however this cannot be the
case of the mean square fluctuations of $K$, in fact out of the sum $K_R$ 
of all the sectional curvatures, in Eq. (\ref{eq_Hill}) only one
sectional curvature is ``picked-up'' from point to point by $\delta K$ 
so that $\delta K$ remains finite with increasing $N$. Therefore, as the second 
cumulant of $\delta K$ does not vanish with $N$, we have to keep finite the
second cumulant of $\delta K_R$, what is simply achieved by properly adjusting 
the numerical factor in Eq.  (\ref{momento2}).

The lowest order approximation of a cumulant expansion of the stochastic 
process $\delta K(s)$ is the gaussian approximation
\begin{equation}
\delta K(s) \simeq
\frac{1}{\sqrt{N-1}}\langle \delta^2 K_R \rangle^{1/2}_s\, \eta(s)~,
\label{K_stocastica}
\end{equation}
where $\eta(s)$ is a random gaussian process with zero mean and unit variance.
Finally, in order to decouple the stability equation from the dynamics, we 
replace time averages with static averages computed with a suitable ergodic
invariant measure $\mu$. As we deal with autonomous Hamiltonian systems, a
natural choice is the microcanonical measure on the constant energy surface
of phase space \cite{nota1}
\begin{equation}
\mu\propto\delta ({\cal H} - E)
\label{misuramicro}
\end{equation}
so that Eq. (\ref{K_stocastica}) becomes
\begin{equation}
\delta K(s) \simeq 
\frac{1}{\sqrt{N-1}}\langle \delta^2 K_R \rangle^{1/2}_\mu\,\eta(s)~.
\label{K_stoc_mu}
\end{equation}
Similarly,  $k_R(s)$ in Eq. (\ref{eq_Hill}) is replaced by 
$\langle k_R\rangle_\mu$, in fact at large $N$ the fluctuations of $k_R$ --
as already noticed above -- vanish as $\frac{1}{N}$ because the coarse-grained
manifold is isotropic, so that we finally have
\begin{equation}
{{d^2\psi}\over{ds^2}}+ \langle k_R\rangle_\mu\, \psi + \frac{1}{\sqrt{N-1}}
\langle\delta^2 K_R\rangle_\mu^{1/2}\,\eta (s)\,\psi = 0~,
\label{eq_Hill_psi}
\end{equation}
where $\psi$ stands for any of the components $J^j$, since all of them
now obey the same effective equation of motion. The instability growth-rate
of $\psi$ measures the instability growth-rate of $\Vert J\Vert^2$ and
thus provides the dynamical instability exponent in our Riemannian framework.
Equation (\ref{eq_Hill_psi}) is a scalar equation which, 
{\em independently of the
knowledge of dynamics}, provides a measure of the average degree of instability
of the dynamics itself through the behavior of $\psi (s)$. The peculiar
properties of a given Hamiltonian system enter Eq. (\ref{eq_Hill_psi}) through
the global geometric properties $\langle k_R\rangle_\mu$ and 
$\langle\delta^2K_R\rangle_\mu$ of the ambient Riemannian manifold (whose
geodesics are natural motions) and are sufficient to determine the average
degree of chaoticity of the dynamics.
Moreover, according to (\ref{misuramicro}),  $\langle k_R\rangle_\mu$ and 
$\langle\delta^2K_R\rangle_\mu$ are functions of the energy $E$ of the system
-- or of the energy density $\varepsilon = E/N$ which is the relevant 
parameter as $N \to \infty$ -- 
so that from (\ref{eq_Hill_psi}) we can obtain the energy dependence of the 
geometric instability exponent.

\subsection{An analytic formula for the largest Lyapunov exponent}
\label{analytform}

By transforming Eq. (\ref{eq_jacobi}) into Eq. (\ref{eq_Hill_psi}) the original 
complexity of the JLC equation has been considerably reduced: from a tensor 
equation we have worked out an effective scalar equation formally representing
a stochastic oscillator. In fact (\ref{eq_Hill_psi}), with a self-evident
notation, is in the form
\begin{equation}
{{d^2\psi}\over{ds^2}}+ \Omega (s)\, \psi =0
\label{eq_stoc_osc}
\end{equation}
where $\Omega (s)$ is a gaussian stochastic process.

Now, passing from proper time $s$ to physical time $t$, Eq. (\ref{eq_stoc_osc})
simply reads
\begin{equation}
{{d^2\psi}\over{dt^2}}+ \Omega (t)\, \psi =0~~,
\label{stoc_osc_t}
\end{equation}
where 
\begin{equation}
\Omega (t)= \langle k_R\rangle_\mu +\frac{1}{\sqrt{N}}
 \langle\delta^2K_R\rangle_\mu^{1/2}\,
\eta (t)
\label{OmegaE}
\end{equation}
if the Eisenhart metric is used [because of the affine parametrization
of the arclength with time, Eq.  (\ref{par_affine})]; if Jacobi metric
is used, we have
\begin{equation}
\Omega (t)= \langle k_R\rangle_\mu + \left\langle -\frac{1}{4}\left(
\frac{\dot W}{W}\right)^2 + \frac{1}{2}\frac{d}{dt}\left(
\frac{\dot W}{W}\right)\right\rangle_\mu +
\frac{1}{\sqrt{N}}\langle\delta^2K_R\rangle_\mu^{1/2}\,\eta (t)
\label{OmegaJ}
\end{equation}
[see Eq.  (64) of \cite{Pettini} and Eq.  (27) of  
\cite{CerrutiPettini1}], note that $d/dt =\dot q^j(\partial /\partial q^j)$. 
Being interested in the large $N$ limit, we replaced $N-1$ with $N$ in Eqs.
(\ref{OmegaE}) and (\ref{OmegaJ}).
Of course Ricci curvature has different
expressions according to the metric used.  

The stochastic process $\Omega (t)$ is not completely determined unless
its time correlation function $\Gamma_\Omega(t_1,t_2)$ is given. 
We consider a stationary and $\delta$-correlated process $\Omega(t)$ so that
$\Gamma_\Omega(t_1,t_2) = \Gamma_\Omega(\vert t_2 - t_1\vert )$ and
\begin{equation}
\Gamma_\Omega(t) = \tau \, \sigma_\Omega^2 \, \delta(t)~, \label{corr_omega}
\end{equation}
where $\tau$ is a characteristic time scale of the process.
In order to estimate $\tau$, let us notice that for a geodesic flow on a smooth
manifold the assumption of  $\delta$-correlation of $\Omega (t)$ will 
be reasonable
only down to some time scale below which the differentiable character of the
geodesics will be felt. In other words, we have to think that in reality the
power spectrum of $\Omega (t)$ is flat up to some high frequency cutoff, 
let us denote it by $\nu_\star$; therefore, by representing the $\delta$ 
function as the limit for $\nu\rightarrow\infty$ of
$\delta_\nu (t)=\frac{\sin (\nu t)}{\pi t}$, a more realistic representation
of the autocorrelation function $\Gamma_\Omega (t)$ in
Eq. (\ref{corr_omega}) could be $\Gamma_\Omega^\star (t)=\sigma^2_\Omega 
\frac{1}{\pi}\frac{\sin (\nu_\star t)}{\nu_\star t}\equiv
\tau_\star\sigma^2
_\Omega \delta_{\nu_\star}(t)$, whence $\tau_\star =1/\nu_\star$.
Notice that $\int_{0^-}^\infty \Gamma_\Omega (t) dt=\tau\sigma_\Omega^2$ and
$\int_{0^-}^\infty \Gamma_\Omega^\star (t) dt=
\frac{1}{2} \tau_\star\sigma_\Omega^2$ thus $\tau =\tau_\star /2$.
For practical computational reasons it is convenient to use 
$\Gamma_\Omega (t)$ in the form given by Eq. (\ref{corr_omega}) 
(with the implicit assumption that $\nu_\star$ is sufficiently large), 
however, being $\nu_\star$ finite, the definition $\tau =\tau_\star /2$ 
will be kept. To estimate $\tau_\star$ we proceed as follows.
A first time scale, which we will refer to as $\tau_1$, is associated  to
the time needed to cover the average distance between
two successive conjugate points along a geodesic \cite{nota2}. In fact, at
distances smaller than this one the geodesics are minimal and far from looking
like random walks, whereas at each crossing of a conjugate point the
separation vector field increases as if the geodesics in the local congruence
were kicked (this is what happens when parametric instability is active).
From Rauch's comparison theorem \cite{DoCarmo} we know that if sectional 
curvature $K$ is bounded as follows: $0< L\leq K \leq H$, 
then the distance $d$ between two
successive conjugate points is bounded by $\frac{\pi}{\sqrt{H}}< d <\frac{\pi}
{\sqrt{L}}$. We need the lower bound estimate that, for strongly convex
domains \cite{Berger}, is slightly modified to $d >\frac{\pi}{2\sqrt{H}}$.

Hence we define $\tau_1$ through
\begin{equation}
\tau_1 = \left\langle\frac{dt}{ds}\right\rangle d_\star =
\left\langle\frac{dt}{ds}\right\rangle \frac{\pi}{2
\sqrt{\Omega_0 +\sigma_{{}_\Omega}}}
\label{dstar}
\end{equation}
where $\left\langle\frac{dt}{ds}\right\rangle$ is the average of the
ratio between proper and physical time 
($\left\langle\frac{dt}{ds}\right\rangle = 1$ if Eisenhart metric is
used) and the upper bound $H$ of $K$ is
replaced by the $N$-th fraction
of a typical peak value of Ricci curvature, which is in turn estimated as its
average $\Omega_0$ plus the typical value $\delta K$ of the (positive)
fluctuation, i.e. in a gaussian approximation $\delta K =
\sigma_{{}_\Omega}$.
This time scale is expected to be the most relevant 
only as long as curvature is positive and the fluctuations, compared to 
the average, are small.

Another time scale, referred to as $\tau_2$, is related to 
local curvature fluctuations. These will be felt on a length scale of 
the order of, at least, $l =1/\sqrt{\sigma^{}_\Omega}$ 
(the average fluctuation of curvature
radius). The scale $l$ is expected to be relevant one when the fluctuations 
are of the same order of magnitude as the average curvature.
When the sectional curvature is positive (resp. negative), 
lengths and time intervals -- on a scale $l$ -- are enlarged
(resp. shortened) by a 
factor $(l^2 K/6)$ 
\cite{nota3}, so that the period $\frac{2\pi}{\sqrt{\Omega_0}}$ 
has a fluctuation amplitude $d_2$ given by 
$d_2 =\frac{l^2 K}{6}\frac{2\pi}{\sqrt{\Omega_0}}$; replacing $K$ 
by its most probable value $\Omega_0$ one gets
\begin{equation}
\tau_2 = \left\langle\frac{dt}{ds}\right\rangle d_2 =
\left\langle\frac{dt}{ds}\right\rangle
 \frac{l^2 \Omega_0}{6}\frac{2\pi}{\sqrt{\Omega_0}}
        \simeq \left\langle\frac{dt}{ds}\right\rangle \frac{\Omega_0^{1/2}}
         {\sigma^{\,}_\Omega}~.
\label{tau2star}
\end{equation}
Finally $\tau$ in Eq. (\ref{corr_omega}) is obtained by combining $\tau_1$
with $\tau_2$ as follows
\begin{equation}
\tau^{-1} = 2\tau_\star^{-1} = 2 \left(\tau_1^{-1} + \tau_2^{-1}\right)~.
\label{taufinale}
\end{equation}
The present estimate of $\tau$ 
is very close -- though not equal -- to the one of Ref. \cite{Lyap}.

Whenever $\Omega (t)$ in Eq. (\ref{stoc_osc_t}) has a non-vanishing stochastic
component the solution $\psi (t)$ has an exponentially growing envelope
\cite{VanKampen} whose growth-rate provides a measure of the degree of
chaoticity. Let us call this quantity Lyapunov exponent
and denote it by $\lambda$.
In the next Section we shall make more precise the relationship of $\lambda$
with the conventional largest Lyapunov exponent.

Our exponent $\lambda$ is defined as
\begin{equation}
\lambda = \lim_{t\to\infty} \frac{1}{2t} \log 
\frac{\psi^2(t) + \dot\psi^2(t)}{\psi^2(0) + \dot\psi^2(0)}~,
\label{def_lambda_gauss}
\end{equation}
where $\psi(t)$ is solution of Eq.  (\ref{stoc_osc_t}).

The ratio $(\psi^2(t) + \dot\psi^2(t))/(\psi^2(0) + \dot\psi^2(0))$ is
computed by means of a technique, developed by Van Kampen and sketched in
Appendix A, which is based on the possibility of computing analytically the
evolution of the second moments of $\psi$ and $\dot\psi$, averaged over the
realizations of the stochastic process, from
\begin{equation}
\frac{d}{dt}\left(
\begin{array}{c}
\langle\psi^2\rangle \\
\langle\dot{\psi}^2\rangle \\
\langle\psi\dot{\psi}\rangle
\end{array} \right) = \left(
\begin{array}{ccc}
0 & 0 & 2 \\
2\sigma^2_\Omega\tau & 0 & -2\Omega_0 \\
-\Omega_0 & 1 & 0
\end{array} \right)\left(
\begin{array}{c}
\langle\psi^2\rangle \\
\langle\dot{\psi}^2\rangle \\
\langle\psi\dot{\psi}\rangle
\end{array} \right)
\label{vankamp}
\end{equation}
where $\Omega_0$ and $\sigma_\Omega$ are respectively the mean and the 
variance of $\Omega (t)$ above defined.
By diagonalizing the matrix in the r.h.s. of (\ref{vankamp}) one finds two
complex conjugate eigenvalues, and one real eigenvalue related to the evolution
of $\frac{1}{2}\left( \langle \psi^2\rangle +\langle\dot\psi^2\rangle\right)$.
According to (\ref{def_lambda_gauss}) the exponent $\lambda$ is half the
real eigenvalue. Simple algebra leads to the final expression
\begin{mathletters}
\begin{equation}
\lambda(\Omega_0,\sigma_\Omega,\tau)  =  \frac{1}{2}
\left(\Lambda-\frac{4\Omega_0}
{3 \Lambda}\right), 
\end{equation}
\begin{equation}
\Lambda  =  \left(2\sigma^2_\Omega\tau+\sqrt{\left(\frac{4\Omega_0}{3}
\right)^3+(2\sigma^2_\Omega\tau)^2}\,\right)^{1/3}.
\end{equation}
\label{Laformula}
\end{mathletters}
All the quantities $\Omega_0$, $\sigma_\Omega$ and $\tau$ can be computed
as {\it static} averages, therefore -- within the validity limits of the 
assumptions made above --  Eqs. (\ref{Laformula}) provide an analytic 
formula to compute
the largest Lyapunov exponent independently of the numerical integration of 
the dynamics and of the tangent dynamics.

\subsubsection{Lyapunov exponent and Eisenhart metric}

Let us consider dynamical systems described by the Lagrangian function
(\ref{eq5}) with a diagonal kinetic energy matrix, i.e. $a_{ij} = \delta_{ij}$,
and let us choose as ambient manifold the enlarged configuration space-time
equipped with the Eisenhart metric (\ref{g_E}).

Trivial algebra gives $\Gamma^i_{00}=(\partial V/\partial q_i)$ and
$\Gamma^{N+1}_{0i}=(-\partial V/\partial q^i)$ as the only nonvanishing
Christoffel coefficients and hence the Riemann curvature tensor has only
the following nonvanishing components
\begin{equation}
R_{0i0j} = \frac{\partial^2 V}{\partial q^i\partial q^j} ~.
\label{riemann_E}
\end{equation}
The JLC equation (\ref{eq_jacobi}) is thus rewritten in local coordinates as
\begin{eqnarray}
{\nabla\over{ds}}{\nabla\over{ds}}J^0 &+&R^0_{i0j}{{dq^i}\over{ds}}J^0
{{dq^j}\over{ds}}+R^0_{0ij}{{dq^0}\over{ds}}J^i{{dq^j}\over{ds}}=0 \nonumber \\
{\nabla\over{ds}}{\nabla\over{ds}}J^i &+&R^i_{0j0}\left( {{dq^0}
\over{ds}}\right) ^2J^j+R^i_{00j}{{dq^0}\over{ds}}J^0{{dq^j}\over{ds}}+
R^i_{j00}{{dq^j}\over{ds}}J^0{{dq^0}\over{ds}}=0 \nonumber \\
{\nabla\over{ds}}{\nabla\over{ds}}J^{N+1}& 
+&R^{N+1}_{i0j}{{dq^i}\over{ds}}J^0
{{dq^j}\over{ds}}+R^{N+1}_{ij0}{{dq^i}\over{ds}}J^j{{dq^0}\over{ds}}=0~. 
\label{JLC_gE}
\end{eqnarray}
As $\Gamma^0_{ij}=0$ implies $\nabla J^0/ds = dJ^0/ds$ and as
$R^0_{ijk}=0$, we find that the first of these equations reads
\begin{equation}
{{d^2J^0}\over{ds^2}}=0~,
\end{equation}
hence $J^0$ does not accelerate and, without loss of generality, we can
set $\dot J^0(0) = J^0(0)=0$, this yields (using $\nabla J^i/ds=dJ^i/ds+
\Gamma^i_{0k}\dot q^0 J^k + \Gamma ^i_{k0}\dot q^k J^0$)
\begin{equation}
{{\nabla^2J^i}\over{ds^2}}={{d^2J^i}\over{ds^2}}
\end{equation}
and the second equation in (\ref{JLC_gE}) gives, 
for the projection in configuration space of the separation vector,
\begin{equation}
\frac{d^2 J^i}{ds^2} + \frac{\partial^2 V}{\partial q_i \partial q_k}
\left(\frac{dq^0}{ds}\right)^2 J_k = 0~~~~~~i=1,...,N; 
\label{eq_jacobi_E}
\end{equation}
the third of equations (\ref{JLC_gE}) describes the passive evolution of
$J^{N+1}$ which does not contribute the norm of $J$ because $g_{N+1N+1}=0$, 
so we can disregard it.

As already mentioned in the previous Section, along the physical geodesics of
$g_E$ it is $ds^2 = (dq^0)^2 = dt^2$ therefore Eq. (\ref{eq_jacobi_E})
is exactly the usual tangent dynamics equation reported in the Introduction,
provided that the obvious identification 
$\xi = (\xi_q,\xi_p)\equiv (J,\dot J)$ is made.
This clarifies the relationship between the geometric description of the
instability of a geodesic flow and the conventional description of dynamical 
instability. It has been recently shown \cite{CerrutiPettini1,ValdetPettini}
that the solutions of the equations (\ref{eq_jacobi_E}) and (\ref{JLCN=2})
(where ${\cal R}$ is computed with Jacobi metric)
are strikingly close one another in the case of two degrees of freedom systems.
This result is reasonable because the geodesics of $(M\times{\Bbb R}^2,g_E)$ 
-- that
are natural motions -- project themselves onto the geodesics of $(M,g_J)$, and 
as the extra coordinates $q^0$ and $q^{N+1}$ do not contribute to the 
instability of the geodesic flow , 
both local and global instability properties must be the same with
either Jacobi or Eisenhart metrics, independently of $N$.

With Eisenhart metric the only nonvanishing component of the Ricci tensor is
$R_{00} = \triangle V$, where $\triangle$ is the euclidean Laplacian in
configuration space.
Hence Ricci curvature is $k_R(q) = \triangle V/(N-1)$  (remember that 
we choose the constant $C$ such that $ds^2 = dt^2$ along a physical geodesic)
and the
stochastic process $\Omega (t)$ in (\ref{stoc_osc_t}) is specified by 
\begin{mathletters}
\begin{equation}
\Omega_0  =  \langle k_R \rangle_\mu  = 
\frac{1}{N}\langle \triangle V \rangle_\mu ~, \label{Omega_0} 
\end{equation}
\begin{equation}
\sigma^2_{{}_\Omega}  =  \frac{1}{N} \langle \delta^2 K_R \rangle_\mu  = 
\frac{1}{N}\left( \langle (\triangle V)^2 \rangle_\mu - 
\langle \triangle V \rangle^2_\mu \right) \label{sigma_Omega}
\end{equation}
\begin{equation}
2 \tau = \frac{\pi\sqrt{\Omega_0}}{2\sqrt{\Omega_0(\Omega_0 +
\sigma_{{}_\Omega})} + \pi\sigma_{{}_\Omega} }~. 
\label{temposcala}  
\end{equation}
\end{mathletters}

\subsubsection{Averages of geometric quantities}
\label{medie.subsec3}
Let us now sketch how to compute the mean and the variance of any
observable function $f(q)$, a geometric quantity of the chosen 
ambient manifold, by means of the microcanonical measure 
(\ref{misuramicro}), i.e.
\begin{equation}
\langle f(q)\rangle_\mu = 
\frac{1}{\omega_{E}}\int f(q)\,\delta({\cal H}(q,p) - E)\,
d q\,d p 
\label{media}
\end{equation}
where 
\begin{equation}
\omega_{E} = \int \delta({\cal H}(q,p) - E)\,d q\,d p
\end{equation}
and $q=(q_1\dots q_N)$, $p=(p_1\dots p_N)$.
By using the configurational partition function $Z_C(\beta )$, given by
\begin{equation}
Z_C(\beta ) =\int d q\,e^{-\beta\,V(q)}
\end{equation}
where $d q=\prod_{i=1}^{N}dq_i$, we can compute
the Gibbsian average $\langle f\rangle^{G}$ of the observable $f$ as 
\begin{equation}
\langle f\rangle^{G} = [Z_{C}(\beta)]^{-1} \int d q\, f(q)\,e^
{- \beta V(q)}~~.
\end{equation}
Whenever this average is known, we can obtain the microcanonical average 
of $f$ \cite{LPV} in the following parametric form
\begin{equation}
\langle f\rangle_{\mu}(\varepsilon) \rightarrow \left\{\begin{array}{l}
\langle f\rangle_{\mu}(\beta) = \langle f \rangle^{G}(\beta)\\
\\
\varepsilon(\beta) = {\displaystyle
\frac{1}{2\beta} - \frac{1}{N}\frac{\partial}{\partial\beta}
[\log Z_{C}(\beta)]}~.
\end{array}\right. 
\label{mm}
\end{equation}
By replacing $f$ with the explicit expression for Ricci curvature
$k_{R} =\frac{1}{N}K_{R}$ we can work out $\Omega_0$.
Notice that Eq. (\ref{mm}) is strictly valid in the thermodynamic limit; 
at finite $N$ it is $\langle f\rangle_{\mu}(\beta) = \langle f \rangle^{G}
(\beta )+ {\cal O}(\frac{1}{N})$.

At variance with the computation of $\langle f\rangle$, which is
insensitive to the choice of the probability measure in the 
$N\rightarrow\infty$ limit, 
computing the fluctuations of $f$, i.e. of 
$\langle \delta^{2}f\rangle=\frac{1}{N}\langle \left(f-\langle f\rangle \right)
^{2}\rangle$, by means of the canonical or microcanonical measures yields 
different results. The relationship between the canonical -- i.e. 
computed with the Gibbsian weight $e^{-\beta{\cal H}}$ -- and the 
microcanonical fluctuations 
is given by the well known formula \cite{LPV}
\begin{equation}
\langle
\delta^{2}f\rangle_\mu (\varepsilon)=\langle\delta^{2}f\rangle^{G}(\beta)-
\frac{\beta^{2}}{C_{V}}
\left[\frac{\partial\langle f\rangle^{G}(\beta)}{\partial\beta}\right]^{2},
\label{corr2}
\end{equation}
where
\begin{equation}
C_{V}=-\frac{\beta^{2}}{N}\frac{\partial\langle E\rangle}{\partial\beta}
\end{equation}
is the specific heat at constant volume and $\beta =\beta (\varepsilon )$ is
given in implicit form by the second equation in (\ref{mm}).

By replacing $f$ with $k_R$ we can work out $\sigma^{2}_\Omega$.

\section{Applications}
\label{sezione4}
In this Section the Riemannian approach to Hamiltonian chaos 
described above is practically used to compute $\lambda_1(\varepsilon )$ for
two different models: the Fermi-Pasta-Ulam (FPU) $\beta$-model and a chain
of coupled rotators.

The choice of these models is motivated by the possibility of analytically 
computing, in the $N\rightarrow\infty$ limit, the geometric quantities 
needed, and by their interest as mentioned in the following subsections.

\subsection {The Fermi-Pasta-Ulam $\beta$-model}
The FPU $\beta$-model is defined by the Hamiltonian \cite{FPU}
\begin{equation}
{\cal H}(p,q)=\sum_{i=1}^N\frac{1}{2}\,p_i^2 + \sum_{i=1}^N\left[
\frac{1}{2}(q_{i+1}-q_i)^2 + \frac{\mu}{4}(q_{i+1}-q_i)^4\right]~.
\label{FPU}
\end{equation}
This is a paradigmatic model of nonlinear classical many-body systems that
has been extensively studied over the last decades and that stimulated
remarkable developments in nonlinear dynamics, one example: the discovery of
solitons. For a recent review we refer to \cite{Ford}.
Also the transition between weak and strong chaos has been first discovered 
in this model \cite{PettiniLandolfi,PettiniCerruti} and then, the effort 
of understanding the origin of such a threshold has stimulated the development
of the geometric theory presented here.

Let us now compute the average Ricci curvature $\Omega_0$ and its fluctuations
$\sigma_\Omega^{\,}$. We have seen above that, using Eisenhart metric, 
$k_R^{\,}$ is given by
\begin{equation}
k_R=\frac{1}{N}\sum_{i=1}^N \frac{\partial^2 V(q)}{\partial q_i^2}~,
\end{equation}
for the FPU $\beta$-model this reads 
\begin{equation}
k_R = 2 + \frac{6\mu}{N} \sum_{i=1}^N \left( q_{i+1} - q_i \right)^2~,
\label{kRFPU}
\end{equation}
note that $k_R^{\,}$ is always positive.

In order to compute the Gibbsian average of $k_R^{\,}$ and its fluctuations,
we rewrite the configurational partition function as
\begin{equation}
\tilde Z_C(\alpha) = \int_{-\infty}^{+\infty} \prod_{i=1}^N dq_i 
\exp\left\{-\beta\sum_{i=1}^N \left[ \frac{\alpha}{2} (q_{i+1}-q_i)^2 +
\frac{\mu}{4} (q_{i+1}-q_i)^4 \right] \right\} ~, \label{Ztilde}
\end{equation}
which, in terms of the arbitrary parameter $\alpha$ and of $Z_C^{\,}$, is
expressed as 
$\tilde Z_C(\alpha) = Z_C \left( \alpha\beta,{\mu}/{\alpha} \right)$
and leads to the following identity
\begin{equation}
\langle k_R \rangle (\beta) = 2 - \frac{12\mu}{\beta N}
\frac{1}{Z_C}\left[ \frac{\partial}{\partial\alpha} \tilde Z_C(\alpha)
\right]_{\alpha=1} ~. \label{kR(Ztilde)}
\end{equation}
Thus we have to compute
\begin{equation}
\frac{1}{NZ_C}\left[ \frac{\partial}{\partial\alpha} \tilde Z_C(\alpha)
\right]_{\alpha=1} = 
\frac{1}{N}\left[ \frac{\partial}{\partial\alpha} \log \tilde Z_C(\alpha)
\right]_{\alpha=1} ~ \label{derlogZtilde}
\end{equation}
using
\begin{equation}
\tilde Z_C(\alpha) = \left[\tilde z_C(\alpha) \right]^N f(\alpha)~,
\end{equation}
where $f(\alpha)$ is a quantity ${\cal O}(1)$, $\tilde z_C(\alpha)$ 
is the single particle partition function \cite{LPRV} 
\begin{equation}
\tilde z_C(\alpha) = \Gamma\left(\frac{1}{2}\right)\left(\frac{\beta\mu}
{2}\right)^{-1/4} \exp (\frac{1}{4}\alpha^2\theta^2) D_{-1/2} (\alpha\theta)~,
\label{ztildeFPU}
\end{equation}
$\Gamma$ is the Euler function, $D_{-1/2}$ is a parabolic cylinder function
and
\begin{equation}
\theta=\left(\frac{\beta}{2\mu}\right)^{1/2}~. \label{theta}
\end{equation}

The final result in parametric form of the average Ricci curvature of
$(M\times{\Bbb R}^2,g_E^{\,})$ -- with the constant energy constraint -- is 
(details can be found in \cite{CasettiPettini}) 
\begin{equation}
\Omega_0 (\varepsilon)\rightarrow \, \left\{ \begin{array}{rcl}
\langle k_R \rangle (\theta) & = & {\displaystyle 
2 + \frac{3}{\theta}\, \frac{D_{-3/2}(\theta)}{D_{-1/2}(\theta)} } \\
& & \\
\varepsilon(\theta) & = & {\displaystyle
\frac{1}{8\sigma}\left[\frac{3}{\theta^2}+\frac{1}{\theta}\,
\frac{D_{-3/2}(\theta)}{D_{-1/2}(\theta)}\right] }
\end{array} \right.
\label{kR(eps)FPU}
\end{equation}
Let us now compute
\begin{equation}
\sigma^2_\Omega (\varepsilon ) = 
\frac{1}{N}\langle \delta^2 K_R \rangle^{\,}_\mu (\varepsilon ) =
\frac{1}{N} \langle \left( K_R - \langle K_R \rangle \right)^2 
\rangle^{\,}_\mu~.
\end{equation}
According to Eq.  (\ref{corr2}), first the Gibbsian average of
this quantity, $\langle \delta^2 k_R \rangle^G (\beta)=
\frac{1}{N} \langle \left( K_R - \langle K_R \rangle \right)^2 \rangle^G
(\beta)$, has to be computed and then the correction term must be added.
Now define 
\begin{equation}
Q=\sum_{i=1}^N (q_{i+1}-q_i)^2 ~;
\end{equation}
after Eq.  (\ref{kRFPU}), 
\begin{equation}
\frac{1}{N}\langle \delta^2 K_R \rangle^G (\beta)=
\frac{1}{N} \langle \left( K_R - \langle K_R \rangle \right)^2 \rangle^G =
\frac{36\mu^2}{N}\langle \left( Q - \langle Q \rangle \right)^2 \rangle^G ~,
\label{fluttQ}
\end{equation}
hence using Eq. (\ref{Ztilde})
\begin{equation}
\langle\left( Q - \langle Q \rangle \right)^2 \rangle^G =
\frac{4}{\beta^2}\left[ \frac{\partial^2}{\partial\alpha^2} 
\log \tilde Z_C(\alpha)\right]_{\alpha=1}~, \label{identFPU2}
\end{equation}
and finally
\begin{equation}
\frac{1}{N}\langle \delta^2 K_R \rangle^G =
\frac{144\mu^2}{\beta^2}\left[ \frac{\partial^2}{\partial\alpha^2} 
\log \tilde z_C(\alpha)\right]_{\alpha=1}~. \label{fluttder2log}
\end{equation}
Simple algebra gives
\begin{equation}
\left[ \frac{\partial^2}{\partial\alpha^2} 
\log \tilde z_C(\alpha)\right]_{\alpha=1} =
\frac{\theta^2}{4} \, \left\{ 2 - 2\theta \, 
\frac{D_{-3/2}(\theta)}{D_{-1/2}(\theta)} -
\left[\frac{D_{-3/2}(\theta)}{D_{-1/2}(\theta)}\right]^2 \right\}~,
\label{der2log}
\end{equation}
so that from Eq. (\ref{fluttder2log}) we obtain
\begin{equation}
\frac{1}{N}\langle \delta^2 K_R \rangle^G (\theta) =
\frac{9}{\theta^2} \, \left\{ 2 - 2\theta \, 
\frac{D_{-3/2}(\theta)}{D_{-1/2}(\theta)} -
\left[\frac{D_{-3/2}(\theta)}{D_{-1/2}(\theta)}\right]^2 \right\}~.
\label{fluttkcan(theta)}
\end{equation}
According to the prescription of Eq.  (\ref{corr2}), the final result for the
fluctuations of Ricci curvature is 
\begin{equation}
\sigma^2_\Omega (\varepsilon ) \rightarrow
\, \left\{ \begin{array}{rcl}
\frac{1}{N}\langle \delta^2 K_R \rangle^{\,}_\mu (\theta) & = & {\displaystyle
\frac{1}{N}\langle \delta^2 K_R \rangle^G (\theta)- \frac{\beta^2}{c_V(\theta)}
\left( \frac{\partial \langle k_R \rangle (\theta)}{\partial\beta}
\right)^2 } \\
 & & \\
\varepsilon(\theta) & = & {\displaystyle
\frac{1}{8\mu}\left[\frac{3}{\theta^2}+\frac{1}{\theta}\,
\frac{D_{-3/2}(\theta)}{D_{-1/2}(\theta)}\right] }
\end{array} \right.
\label{fluttkR(eps)FPU}
\end{equation}
where $\langle \delta^2 K_R \rangle^G (\theta)$ is given by
(\ref{fluttkcan(theta)}), the derivative part of the correction term is
\begin{equation}
\frac{\partial \langle k_R \rangle (\theta)}{\partial\beta} =
\frac{3}{8\mu\theta^3}\, \frac{\theta \, D^2_{-3/2}(\theta)+ 
2(\theta^2 -1)D_{-1/2}(\theta) D_{-3/2}(\theta) -2\theta D^2_{-1/2}(\theta)}
{D^2_{-1/2}(\theta)} ~, \label{dkRdbeta}
\end{equation}
and the specific heat per particle $c^{\,}_{{}_V}$ is found to be
\begin{eqnarray}
c^{\,}_{{}_V}(\theta )&=&
{\displaystyle \frac{1}{16 D^2_{-1/2}(\theta )} }\left\{ (12 + 2\theta^2)
D^2_{-1/2}(\theta ) + 2\theta D_{-1/2}(\theta )D_{-3/2}(\theta )\right.
 \nonumber \\
&-&\left. \theta^2 D_{-3/2}(\theta ) \left[ 2\theta D_{-1/2}(\theta ) +
D_{-3/2}(\theta )\right] \right\}~.
\label{calspec}
\end{eqnarray}
The microcanonical averages in Eqs.(\ref{kR(eps)FPU}) and 
(\ref{fluttkR(eps)FPU}) are compared in Figs. \ref{fig_curv_fpu} 
and \ref{fig_flutt_fpu} 
with their corresponding time averages computed along numerical trajectories 
of the model (\ref{FPU}) at $N=128$ and $N=512$ with $\mu = 0.1$. The
equations of motion
are integrated using a third order bilateral symplectic algorithm \cite{Lapo}
which is a high precision numerical scheme.

Though microcanonical averages are computed in the
thermodynamic limit, the agreement between time and ensemble averages 
is excellent already at $N=128$.

\subsubsection {Analytic result for $\lambda_1(\varepsilon)$ and its comparison
with numeric results}

Now we use (\ref{kR(eps)FPU}) and (\ref{fluttkR(eps)FPU}) to compute 
$\tau$ according to its definition in (\ref{temposcala}), then we substitute 
$\Omega_0(\varepsilon )$, 
$\sigma_\Omega^2(\varepsilon )$ and $\tau (\varepsilon )$ into Eq. (\ref{Laformula}) 
to obtain the analytic prediction for $\lambda_1(\varepsilon )$ in the limit
$N\rightarrow\infty$.
In Fig. \ref{fig_lyap_fpu} 
this analytic result is compared to the numeric values of
$\lambda_1$ computed by means of the standard algorithm \cite{BGS} at
$N=256$ and $N=2000$ with $\mu = 0.1$ and at different $\varepsilon$.
The agreement between analytic and numeric results is strikingly good.

\subsection{A chain of coupled rotators}
\label{modgaussrot}

Let us now consider the system described by the Hamiltonian
\begin{equation}
{\cal H}({p},{q})
=\sum_{i=1}^{N}\left\{\frac{p_{i}^{2}}{2}+J[1-\cos(q_{i+1}-q_{i})]\right\}.
\label{hamiltonianarot1d}
\end{equation}
If the canonical coordinates $q_i$ and $p_i$ are given the meaning of angular
coordinates and momenta, this Hamiltonian describes a linear chain
of $N$ rotators constrained to rotate on a plane and coupled by a
nearest-neighbor interaction.

This model can be formally obtained by restricting to one spatial dimension 
the classical Heisenberg model whose potential energy is 
$V= -J\sum_{\langle i,j \rangle}{\bf S}_{i}\cdot{\bf S}_{j}$, where the sum is
extended only over nearest-neighbor pairs, $J$ is the coupling constant and
each ${\bf S}_{i}$ has unit module and rotates on a plane.
To each ``spin'' ${\bf S}_{i}= (\cos q_{i},\sin q_{i})$ the velocity 
$\frac{d}{dt}{\bf S}_{i}=(-\frac{d q_{i}}{dt} \sin q_{i},
\frac{dq_{i}}{dt} \cos q_{i})$ is associated so that (\ref{hamiltonianarot1d})
follows from ${\cal H}=\sum_{i=1}^{N}
\frac{1}{2}\dot{\bf S}_{i}^{2} -J \sum_{\langle i,j \rangle} {\bf
S}_{i}\cdot {\bf S}_{j}$.

The Hamiltonian (\ref{hamiltonianarot1d}) has two integrable limits. In the 
limit of vanishing energy it represents a chain of harmonic oscillators
\begin{equation}
{\cal H}({p},{q})
\simeq\sum_{i=1}^{N}\left\{\frac{p_{i}^{2}}{2}+J
(q_{i+1}-q_{i})^2\right\}~,
\label{hamrotE0}
\end{equation}
whereas in the limit of indefinitely growing energy a system of freely rotating
objects is found because of potential boundedness.

The expression of Ricci curvature $K_{R}$, computed with Eisenhart metric, is
\begin{equation}
K_{R} = \sum_{i=1}^{N}\frac{\partial^{2}V({q})}{\partial \,q_{i}^{2}}
=2J\sum_{i=1}^{N}\cos(q_{i+1}-q_{i}).
\end{equation}
Let us observe that for this model a relation exists between potential energy
$V$ and Ricci curvature $K_R$:
\begin{equation}
V({q})= JN - \frac{K_{R}}{2}.
\label{vincolo}
\end{equation}
This relation binds the fluctuating quantity that enters the analytic formula
for $\lambda_1$. This constraint does not exist for  the sectional curvature
thus a-priori it may be expected that some problem will arise.

The configurational partition function for a chain of coupled rotators is
\begin{eqnarray}
{Z}_{C}(\beta) 
&=& \int_{-\pi}^{\pi}\prod_{i=1}^{N}dq_{i} \exp\left\{-\beta\sum_{i=1}^{N}
J[1-\cos(q_{i+1}-q_{i})]\right\} \nonumber \\
&=& \exp (-\beta JN)
\int_{-\pi}^{\pi}\prod_{i=1}^{N}d\omega_{i} 
 \exp(\beta J\sum_{i=1}^{N}\cos\omega_{i}) \label{Z} \\
&=& \exp (-\beta \,JN)[{I}_{0}(\beta \,J)]^{N}(2\pi)^{N}g(\overline{\omega})~.
\nonumber 
\end{eqnarray}
where ${I}_{0}(x) = \frac{1}{\pi}\int _{0}^{+\pi}e^{x\cos\theta}d\theta$ is the
modified Bessel function of index zero; $\omega_i=q_{i+1}-q_i$,
$i\in (1,\dots,N-1)$, $\omega_N={\overline q}-q_N$, ${\overline q}=
{\overline\omega}$ depend on the initial conditions. The function
$g({\overline\omega})$ contributes with a term of ${\cal O}(\frac{1}{N})$ thus
vanishing in the thermodynamic limit.

In order to compute $\Omega_0$ and $\sigma^2_\Omega$ we follow the same
procedure adopted for the FPU model, i.e. we define 
\begin{eqnarray}
\tilde{Z}_{C}(\alpha) &=& \int_{-\pi}^{+\pi}\prod_{i=1}^{N}dq_{i}
\exp\left\{-\beta\sum_{i=1}^{N}[1-\alpha\cos(q_{i+1}-q_{i})]\right\}\nonumber\\
&=& \exp (-\beta JN)\left[I_{0}(\beta J
\alpha)\right]^{N}g(\overline{\omega})(2\pi)^{N}
\end{eqnarray}
and by observing that
\begin{equation}
\langle k_{R}\rangle_{{}_\mu}(\beta)= 
\frac{2}{N\beta}\left[\frac{\partial}{\partial\alpha}
\log \tilde{Z}_{C}(\alpha)\right]_{\alpha=1}.
\label{dlogZ}
\end{equation}
we find $\Omega_0(\varepsilon )$ in parametric form
\begin{equation}
\Omega_0(\varepsilon) \rightarrow \left\{\begin{array}{l}
\langle k^{\,}_{R}\rangle_{{}_\mu}(\beta)= 
2J\,{\displaystyle\frac{I_{0}(\beta J)}
{I_{1}(\beta J)}} \\
\\
\varepsilon(\beta)={\displaystyle\frac{1}{2\beta} }+
J\left(1- \,{\displaystyle\frac{I_{1}(\beta J)}{I_{0}(\beta J)} }\right).
\end{array}\right.
\label{media-rot}
\end{equation}
In order to work out the average of the square fluctuations of Ricci curvature
we use the following identity
\begin{equation}
\frac{1}{N}\langle \delta^{2}K_{R}\rangle^{G}=
\frac{4}{\beta^{2}N}\left[\frac{\partial^{2}}{\partial
\alpha^{2}}\log\tilde{Z}_{C}(\alpha)\right]_{\alpha=1}
\end{equation}
whence
\begin{equation}
\frac{1}{N}\langle \delta^{2}K_{R}\rangle^{G}=4J^{2}\frac{\beta J I_{0}^{2}
(\beta J)-I_{1}(\beta J)I_{0}(\beta J)-\beta J I_{1}^{2}(\beta J)}{\beta J 
I_{0}^{2}(\beta J)}.
\end{equation}
The computation of the correction term $\left[\frac{\partial\langle
k_{R}\rangle(\beta)}{\partial\beta}\right]^{2}/\frac{\partial\varepsilon(\beta)}
{\partial \beta}$ involves the following derivatives
\begin{equation}
\frac{\partial\varepsilon(\beta)}{\partial\beta}= - \frac{1}{2\beta}-
J^{2}\left\{1-
\frac{1}{\beta J}\frac{I_{1}(\beta J)}{I_{0}(\beta J)}-\left[\frac{I_{1}(\beta
J)}{I_{0}(\beta J)}\right]^{2}\right\}
\end{equation}
\begin{equation}
\frac{\partial \langle k_{R}\rangle(\beta)}{\partial\beta}=
2 J^{2}\left\{1-
\frac{1}{\beta J}\frac{I_{1}(\beta J)}{I_{0}(\beta J)}-\left[\frac{I_{1}(\beta
J)}{I_{0}(\beta J)}\right]^{2}\right\}~.
\end{equation}
Finally, gluing together the different terms, we obtain
\begin{equation}
\sigma^2_\Omega (\varepsilon) \rightarrow \left\{\begin{array}{l}
\frac{1}{N}\langle\delta^{2}K_{R}\rangle(\beta)={\displaystyle 
\frac{ 4 J}{\beta}\frac{\beta J I_{0}^{2}(\beta
J)-I_{0}(\beta J)I_{1}(\beta J)-\beta J I_{1}^{2}(\beta J)}
{I_{0}^{2}(\beta
J)\left[1+2\left(\beta J\right)^{2}\right]-2\beta J I_{1}(\beta J)I_{0}(\beta
J)-2\left[\beta J I_{1}(\beta J)\right]^{2}} }\\
\\
\varepsilon(\beta)={\displaystyle \frac{1}{2\beta}+J\left[1-
\frac{I_{1}(\beta J)}{I_{0}(\beta J)}\right]}.
\end{array}\right.
\label{flutt-rot}
\end{equation}
In Figs. \ref{fig_curv_rot} and \ref{fig_flutt_rot} 
the comparison between analytic and numeric results is
provided for the average Ricci curvature and its fluctuations. The agreement 
between ensemble and time averages is very good. Time averages
are computed along numerical trajectories of the model Hamiltonian 
(\ref{hamiltonianarot1d})
at $N=150$ and $J=1$. The already mentioned high precision symplectic algorithm
has been used also in this case.
\subsubsection{Analytic result for $\lambda_1(\varepsilon)$ and its comparison 
with numeric results}

By inserting into Eq. (\ref{Laformula}) the analytic expressions of 
$\Omega_0(\varepsilon)$ and $\sigma^2_\Omega (\varepsilon)$ given in Eqs. 
(\ref{media-rot}) and (\ref{flutt-rot}) -- and also
$\tau(\varepsilon)$ which is a function of the latter quantities -- 
we find $\lambda_1(\varepsilon)$.

In Fig. \ref{fig_lyap_rot} the comparison is given 
between the analytic result so obtained and the outcome of 
numeric computations performed with the standard algorithm \cite{BGS}.
Figure \ref{fig_lyap_rot} shows that there is agreement between analytic 
and numeric values of the largest Lyapunov exponent only at low and high
energy densities. Likewise the FPU case, at low energy, in the quasi-harmonic
limit, we find $\lambda_1(\varepsilon)\propto\varepsilon^2$. Whereas at high energy
$\lambda_1(\varepsilon)\propto\varepsilon^{-1/6}$, here $\lambda_1(\varepsilon)$ is a 
decreasing function because at $\varepsilon\rightarrow\infty$ the
systems is integrable.
In an intermediate energy range our theoretical prediction
underestimates the actual degree of chaoticity of the system.
It is worth mentioning that this energy range coincides with a region of fully
developed (strong) chaos detected in this model by a completely
different approach in Ref. \cite{EKLR}. In this case -- as already mentioned 
above -- there was a-priori a reason to expect an inadequacy of the
analytic prediction in some energy range. In fact, using Eisenhart metric, 
the explicit
expression of the sectional curvature $K({v},{\xi})$ -- relative to 
the plane spanned by
the velocity vector ${v}$ and a generic vector ${\xi} \bot {v}$
(here we use $\xi$ to denote the geodesic separation vector in order to avoid
confusion with $J$ which is the notation for the coupling constant) -- is
\begin{equation}
K({v}, \xi)=R_{0i0k}\frac{dq^0}{dt}\frac{\xi^i}{\Vert{\xi}\Vert}
\frac{dq^0}{dt}\frac{\xi^k}{\Vert{\xi}\Vert} \equiv 
\frac{\partial^{2}V}{\partial q^{i}\partial q^{k}}\frac{\xi^i \xi^k}
{\Vert{\xi}\Vert^{2}}~,
\end{equation}
hence we get 
\begin{equation}
K({v},{\xi})=\frac{J}{\Vert{\xi}\Vert^{2}}\sum_{i=1}^{N}
\cos(q_{i+1}-q_{i})\left[\xi^{i+1}-\xi^{i}\right]^{2}
\label{csezrot}
\end{equation}
for the coupled rotators model.
We realize, by simple inspection of Eq.  (\ref{csezrot}), 
that $K$ can take negative values with non-vanishing probability regardless 
of the value of $\varepsilon$, 
whereas -- as long as $\varepsilon < J$ -- 
this possibility is lost in the replacement of $K$ by Ricci 
curvature that we adopted in our theory. In fact, because of the constraint
(\ref{vincolo}), at each point of the manifold it is
\begin{equation}
k_R(\varepsilon )\geq 2(J-\varepsilon)
\end{equation}
thus our approximation fails in accounting for the
presence of negative sectional curvatures at small values of $\varepsilon$. 
In Eq.  (\ref{csezrot}) the cosines have different and variable weights, 
$[\xi^{i+1}-\xi^i]^2$, that in principle make possible to find somewhere
along a geodesic $K<0$ also with only one negative cosine. This is not the
case of $k_{{}_R}$ where all the cosines have the same weight. Therefore
the probability of finding $K<0$ along a geodesic must be related to the
probability of finding an angular difference greater than $\frac{\pi}{2}$ 
between two nearest-neighboring rotators.
If the energy is sufficiently low this event will be very unlikely, but we
can guess that it will become considerable where the theoretical prediction
is not satisfactory, i.e. when chaos is strong.
Notice that the frequent occurrence of $K<0$ along a geodesic adds to 
parametric instability another instability mechanism that enforces chaos
[Eq. (\ref{cosh})].

Our strategy is to modify the model for $K(s)$ in some {\it effective} 
way that takes into account the mentioned difficulty of $k_R(s)$ to
adequately model $K(s)$. This will be achieved by suitably ``renormalizing'' 
$\Omega_0$ or $\sigma_{{}_\Omega}$ to obtain an {\it effective gaussian 
process} for the behavior of the sectional curvature.

From Eq. (\ref{csezrot}) we see that $N$ directions of the vector $\xi$ exist
such that the sectional curvatures -- relative to the $N$ planes spanned by 
these vectors together with $v$ -- are just $\cos (q_{i+1}- q_i)$. Hence
the probability $P(\varepsilon )$ of occurrence of a negative value of the
cosine is used to estimate the probability of occurrence of negative sectional
curvatures along the geodesics. This probability function has the following
simple expression
\begin{equation}
P(\varepsilon) = \frac{\int_{-\pi}^{\pi}\Theta(-\cos x)e^{\beta J \cos
x}dx}{\int_{-\pi}^{\pi}e^{\beta J \cos x}dx}=
\frac{\int_{\frac{\pi}{2}}^{\frac{3\pi}{2}}e^{\beta J \cos
x}dx}{2\pi I_{0}(\beta J)},
\label{prob.cos.neg}
\end{equation}
where $\Theta(x)$ is the Heavyside unit step function. 

The function $P(\varepsilon)$, reported in Fig. \ref{fig_pro_neg}, begins to
increase at $\varepsilon\simeq 0.2$, just where the analytic prediction in
Fig. \ref{fig_lyap_rot} 
begins to fail, and when it approaches its asymptotic value of 
$\frac{1}{2}$, around the end of the knee, a good agreement is again found 
between theory and numeric results. The simplest way to account for the 
existence of negative sectional curvatures is to shift the peak of the 
distribution ${\cal P}(\delta K_R)$ toward the negative axis. This is
achieved by the replacement
\begin{equation}
\langle k_{R}(\varepsilon)\rangle\rightarrow 
\frac{\langle k_{R}(\varepsilon)\rangle} {1+\alpha P(\varepsilon)}.
\label{Kcorr}
\end{equation}
This correction neither has influence when $P(\varepsilon)\simeq 0$ (below 
$\varepsilon\simeq 0.2$) nor when $P(\varepsilon)\simeq 1/2$ (because in this case
$\langle k_{R}(\varepsilon)\rangle\rightarrow 0$).
The value of the parameter $\alpha$ in (\ref{Kcorr}) must be estimated 
{\em a posteriori} in order to obtain the best agreement between numerical
and theoretical data over the whole range of energies. The result shown
in Fig. \ref{fig_lyap_rot_corr} is obtained with $\alpha = 150$,
anyhow no particularly fine tuning is necessary to obtain a very good
agreement between theory and numerical experiment. 

\section{Concluding remarks}
\label{sezione5}

The present paper contains a substantial progress along the research line
initiated in Ref. \cite{Pettini} where it was proposed to tackle Hamiltonian 
chaos using the Riemannian geometrization of newtonian dynamics.
This work renewed an old intuition that dates back to N. S. Krylov 
\cite{Krylov} and that spawned new ideas in abstract ergodic 
theory \cite{Sinai,Anosov}, 
whereas it did not give rise to any useful method to describe chaos in 
physical geodesic flows, despite of many attempts and with remarkable 
exceptions \cite{Knauf,Gutzwiller}. The obstacle was always the same: in analogy with 
Anosov flows, that live on hyperbolic manifolds, chaos has been invariably 
thought of as a consequence only of negative scalar curvature.
So the first obvious check against any typical model that undergoes a
stochastic transition -- say the H\'enon-Heiles model -- gives a puzzling
surprise: the scalar curvature of $(M,g_{{}_J})$ is always positive
\cite{CerrutiPettini1} independently of the energy value, i.e. of regular
or chaotic behavior of the dynamics.

The novelty of the approach started in Ref. \cite{Pettini} was to conjugate
theoretical arguments with numerical experiments in order to shine some light
on the following two points: {\it i)} does the geometry of the ``mechanical''
manifolds contain, though in some hidden way, the relevant information 
concerning stability and instability of their geodesics? and in the affirmative
case {\it ii)} how to quantify the strenght of chaos, how to characterize 
the weakly and strongly chaotic regimes?

Actually positive answers to these questions have been given in 
\cite{Pettini,CasettiPettini,Lyap,CerrutiPettini,CerrutiPettini1,ValdetPettini}, 
where,
among other things, it has been shown that if the geodesics feel a positive 
non-constant curvature of the underlying manifold then {\it parametric
instability} can be activated. Though a rigorous proof is not yet at disposal, 
parametric instability  appears as the source of chaos on manifolds of 
positive non-constant curvature.

By the way we can mention that also in the case of integrable systems, whose
geodesics are therefore stable, the curvature of the underlying manifold can
be wildly fluctuating along the geodesics but in this case the parametric
instability mechanism is inactive, and it is found that these integrable 
geodesic flows have very special hidden symmetries, mathematically defined
through Killing tensor fields \cite{clementi}, that make them peculiar.

For geodesic flows on constant negative curvature manifolds, the instability 
exponent is known [Eq. (\ref{cosh})], if the curvature is negative and 
non-constant then simple
averaging algorithms can be devised, but what can we do with a positive
and fluctuating curvature? The challenge was now to compute the average
instability exponent for geodesic flows of physical relevance.
This is a crucial test of effectiveness of the Riemannian theory of chaos
with respect to the conventional explanation based on homoclinic intersections.
Moreover, as no analytic method was available to compute Lyapunov exponents,
it was worth making an effort in this direction.

Under reasonable hypotheses, that obviously restrict the domain of validity
of the analytic formula (\ref{Laformula}) for $\lambda_1$, this paper 
provides the first analytic computations of the largest Lyapunov exponent
in dynamical systems described by ordinary differential equations.

Though several points need a deeper understanding, 
we hope that our work convincingly shows that this geometric approach is
effective and useful, thus deserving  further improvements and developments.

\acknowledgments
During the long preparation of this paper we have profited 
of several discussions
with S. Caracciolo, R. Livi, M. Rasetti and G. Vezzosi.
A particularly warm acknowledgment is addressed to L. Caiani for his helpful
criticism.

The final stage of this work was partially supported by ISI Foundation and
by EU HC\&M Network ERBCHRX-CT940546.

\appendix
\section{Solution of the stochastic oscillator equation}

In the following we will briefly describe how to cope with the stochastic 
oscillator problem which we encountered in Sec. \ref{analytform}. The 
discussion follows closely Van Kampen (Ref. \cite{VanKampen}) where all
the details can be found.

A stochastic differential equation can be put in the general form
\begin{equation}
F(x,\Omega)=0,
\label{a1}
\end{equation}
where $F$ is an assigned function and the variable $\Omega$ is a random
process, defined by a mean, a standard deviation and an autocorrelation
function. A function $~{\xi}(\Omega)$ is a solution of this equation if
$\forall \Omega$ $F(\xi (\Omega ),\Omega)=0$.
If equation (\ref{a1}) is linear of order $n$, it is written as
\begin{equation}
\dot{{\bf u}}={\bf A}(t,\Omega){\bf u}
\label{stoclin}
\end{equation}
where ${\bf u}\in {\Bbb R}^{n}$ and ${\bf A}$ is a $n\times n$ matrix whose 
elements are randomly dependent on time.

For the purposes of our work we are interested in studying 
the evolution of the 
average carried over all the realizations of the process , 
$\langle {\bf u}(t)\rangle$. Let us consider the matrix ${\bf A}$ as the sum
\begin{equation}
{\bf A}(t,\Omega)={\bf A}_{0}(t)+\alpha{\bf A}_{1}(t,\Omega)
\label{decomp}
\end{equation}
where the first term is $\Omega$-independent and the second one is randomly
fluctuating with zero mean.
Let us also assume that ${\bf A}_{0}$ is time-independent.
If the parameter $\alpha$ -- that determines the fluctuation amplitude --
is small we can treat Eq. (\ref{stoclin}) by means of a perturbation expansion.
It is convenient to use the interaction picture, thus we put
\begin{equation}
{\bf u}(t)=\exp({\bf A}_{0}t){\bf v}(t)
\end{equation}
\begin{equation}
{\bf A}_{1}(t)=\exp({\bf A}_{0}t){\bf v}(t)\exp(-{\bf A}_{0}t).
\end{equation}
Formally one is led to a Dyson expansion for the solution ${\bf v}(t)$. 
Then, going back to the previous variables and averaging, the second order
approximation gives
\begin{equation}
\frac{d}{dt}\langle{\bf u}(t)\rangle=\{{\bf
A}_{0}+\alpha^{2}\int^{+\infty}_{-\infty}\langle{\bf A}_{1}(t)\exp({\bf
A}_{0}\tau){\bf A}_{1}(t-\tau)\rangle\exp(-{\bf A}_{0}\tau)d\tau\}\langle{\bf
 u}(t)\rangle~~~.
\label{media2}
\end{equation}
Following the same procedure one can find also the evolution of the second
moments (and by iterating also the evolution of higher moments).
In fact, with the components of ${\bf u}\in {\Bbb R}^{n}$ we can make
$n^{2}$ quantities $u_{\nu}u_{\mu}$ that obey the differential equation
\begin{equation}
\frac{d}{dt}(u_{\nu}u_{\mu})=\sum_{k,\lambda}\tilde{A}_{\nu\mu,k\lambda}(t)(
u_{k}u_{\lambda}),
\end{equation}
where
\begin{equation}
\tilde{A}_{\nu\mu,k\lambda}=A_{\nu
k}\delta_{\mu\lambda}+\delta_{\nu k}A_{\mu \lambda}~~~.
\label{Atilde}
\end{equation}
The above presented averaging method can be now applied to this new equation.

Now, if we consider a random harmonic oscillator, Eq. (\ref{stoclin}) has the 
form
\begin{equation}
\frac{d}{dt}\left(
\begin{array}{c}
x \\
\dot{x}
\end{array}\right)=\left(\begin{array}{cc}
0 & 1 \\
-\Omega & 0 \end{array}\right)\left(\begin{array}{c}x \\
\dot{x}\end{array}\right),
\end{equation}
with the random squared 
frequency $\Omega =\Omega_{0}+\sigma^{\,}_{\Omega}\eta(t)$.
In particular, we are interested in working out the second moments equation
when the process $\eta(t)$ is gaussian and $\delta$-correlated.
Using Eq. (\ref{Atilde}) one finds that
\begin{equation}
\frac{d}{dt}\left(\begin{array}{c}
x^{2} \\ \dot{x}^{2} \\ x\dot{x} \end{array} \right) = \left( \begin{array}{ccc}
0 & 0 & 2 \\
0 & 0 & -2\Omega \\
-\Omega & 1 & 0  \end{array} \right)\left(\begin{array}{c} x^{2} \\
\dot{x}^{2} \\ x\dot{x} \end{array} \right)
={\bf A}
\left(\begin{array}{c} x^{2} \\
\dot{x}^{2} \\ x\dot{x} \end{array} \right)~~~.
\end{equation}
Because of our assumptions for this system, Eq. (\ref{media2}) is more than a
second order approximation, it is exact.
In fact, the Dyson series can be written in compact form as
\begin{equation}
\left(\begin{array}{c}
\langle x^{2}(t)\rangle \\ \langle\dot{x}^{2}(t)\rangle \\ 
\langle x(t)\dot{x}(t)\rangle
\end{array}\right)=\lceil\langle\exp\left(\int_{0}^{t}{\bf A}(t')dt'\right)
\rangle
\rceil\left(\begin{array}{c}
\langle x^{2}(0)\rangle \\ \langle\dot{x}^{2}(0)\rangle \\ \langle x(0)
\dot{x}(0)\rangle
\end{array}\right)
\label{tord}
\end{equation}
where the brackets $\lceil\ldots\rceil$ stand for a chronological product.
According to Wick's procedure we can rewrite Eq.  (\ref{tord}) as a cumulant
expansion, and 
when the cumulants of order higher than the second vanish (as is the case of 
interest to us) one can easily show that
Eq.  (\ref{media2}) is exact.

Likewise in Eq.  (\ref{decomp}), the matrix ${\bf A}$ splits as
\begin{equation}
{\bf A}(t)={\bf A}_{0}+\sigma^{\,}_{\Omega}\eta(t){\bf A}_{1}=
\left(\begin{array}{ccc}
0 & 0 & 2 \\
0 & 0 & -2\Omega_{0} \\
-\Omega_{0} & 1 & 0
\end{array}\right)+\sigma^{\,}_{\Omega}\eta(t)\left(\begin{array}{ccc}
0 & 0 & 0 \\
0 & 0 & -2 \\
-1 & 0 & 0 \end{array}\right)
\end{equation}
therefore the equation for the averages becomes
\begin{equation}
\frac{d}{dt}
\left(\begin{array}{c}
\langle x^{2}\rangle \\ \langle\dot{x}^{2}\rangle \\ \langle x\dot{x}\rangle
\end{array}\right)=\{{\bf
A}_{0}+\sigma^{2}_{\Omega}\int_{-\infty}^{+\infty}\langle\eta(t)\eta(t-\tau)
\rangle{\bf B}(\tau)d\tau
\}
\left(\begin{array}{c}
\langle x^{2}\rangle \\ \langle\dot{x}^{2}\rangle \\ \langle x\dot{x}\rangle
\end{array}\right),
\end{equation}
where ${\bf B}(\tau)={\bf A}_{1}\exp({\bf A}_{0}\tau){\bf A}_{1}
\exp(-{\bf A}_{0}\tau)$.
As $\langle\eta(t)\eta(t-\tau)\rangle=\tau \delta(\tau)$, with
$\tau$ a characteristic time scale of the process, we obtain
\begin{equation}
\frac{d}{dt}
\left(\begin{array}{c}
\langle x^{2}\rangle \\ \langle\dot{x}^{2}\rangle \\ \langle x\dot{x}\rangle
\end{array}\right)=\{{\bf
A}_{0}+\sigma^{2}_{\Omega}\tau {\bf B}(0)
\}
\left(\begin{array}{c}
\langle x^{2}\rangle \\ \langle\dot{x}^{2}\rangle \\ \langle x\dot{x}\rangle
\end{array}\right)~~~.
\end{equation}
From the definition of ${\bf B}(\tau)$ it follows that 
${\bf B}(0)={\bf A}_{1}^{2}$, then by easy calculations we find 
\begin{equation}
{\bf A}_{0}+\sigma^{2}_{\Omega}\tau {\bf A}_{1}^{2}=\left(\begin{array}{ccc}
0 & 0 & 2 \\
2\sigma^{2}_{\Omega}\tau & 0 & -2\Omega_{0} \\
-\Omega_{0} & 1 & 0 \end{array}\right)
\end{equation}
which is the result used in Sec. \ref{analytform}.

\begin{figure}

\caption{Average Ricci curvature $\langle k_R \rangle$ vs. energy density
$\varepsilon$ for the FPU model: comparison between analytic
computation
with Eq. (\protect\ref{kR(eps)FPU}) (solid line)  and the outcome
of numerical simulations (time averages) with $N=128$ (solid circles)
and $N=512$ (solid triangles); $\mu = 0.1$. 
\label{fig_curv_fpu}}
\end{figure}

\begin{figure}

\caption{Fluctuation of  Ricci curvature $\langle \delta^2 K_R \rangle/N$ 
vs. energy density $\varepsilon$ 
for the FPU model: comparison between analytic computation
with Eq. (\protect\ref{fluttkR(eps)FPU}) (solid line)  and numerical 
results. Symbols and parameters as in Fig. \protect\ref{fig_curv_fpu}.
\label{fig_flutt_fpu}}
\end{figure}

\begin{figure}

\caption{Lyapunov exponent $\lambda_1$ vs. energy density
$\varepsilon$ for the FPU model: comparison between theoretical prediction
of Eq. (\protect\ref{Laformula}) (solid line) and numerical estimates at
$N = 256$ (solid circles) and $N=2000$ (solid squares); $\mu = 0.1$.
\label{fig_lyap_fpu}}
\end{figure}

\begin{figure}

\caption{Average Ricci curvature $\langle k_R \rangle$ vs. energy density
$\varepsilon$ for the coupled rotators model: 
comparison between analytic computation
with Eq. (\protect\ref{media-rot}) (solid line)  and the outcome
of numerical simulations (time averages) with $N=150$ (solid circles); $J = 1$.
\label{fig_curv_rot}}
\end{figure}

\begin{figure}

\caption{Fluctuation of  Ricci curvature $\langle \delta^2 K_R \rangle/N$ 
vs. energy density $\varepsilon$ 
for the FPU model: comparison between analytic computation
with Eq. (\protect\ref{flutt-rot}) (solid line)  and numerical 
results. Symbols and parameters as in Fig. \protect\ref{fig_curv_rot}.
\label{fig_flutt_rot}}
\end{figure}

\begin{figure}

\caption{Lyapunov exponent $\lambda_1$ vs. energy density
$\varepsilon$ for the coupled rotators model: 
comparison between theoretical prediction
of Eq. (\protect\ref{Laformula}) (solid line) and numerical estimates at
$N = 150$ (solid circles), $N=1000$ (solid rhombs) and $N=1500$ (solid square);
 $J = 1$. 
\label{fig_lyap_rot}}
\end{figure}

\begin{figure}

\caption{Estimate of the probability $P(\varepsilon)$ 
of occurrence of negative sectional curvatures in the coupled rotators model
according to Eq. (\protect\ref{prob.cos.neg}); $J = 1$.
\label{fig_pro_neg}}
\end{figure}

\begin{figure}

\caption{Lyapunov exponent $\lambda_1$ vs. energy density
$\varepsilon$ for the coupled rotators model: 
comparison between theoretical prediction
and numerical estimates as
in Fig. \protect\ref{fig_lyap_rot}, but here the average curvature 
$\langle k_R \rangle$ which enters Eq. (\protect\ref{Laformula})  
is corrected according to Eq. (\protect\ref{Kcorr})
with $\alpha = 150$.  Numerical values of $\lambda_1$ are obtained at
$N=150$ (solid circles), at $N=1000$ (solid rhombs) and at $N=1500$ (solid
square).
\label{fig_lyap_rot_corr}}
\end{figure}


\begin{references}
\bibitem{nota1} At $N\geq 3$, after the Poincar\'e-Fermi theorem, generic
nonintegrable systems have no smooth invariant besides energy, thus the whole
constant energy surface is topologically accessible. One might think that
troubles with ergodicity could be raised by the KAM theorem, but at large $N$
this is not possible because a positive measure of tori survives a perturbation
of amplitude smaller than some threshold value, and the threshold typically
drops to zero as $\exp(- c N\ln N)$, $c$ is a constant.
\bibitem{nota2} The conjugate points are defined by the vanishing of Jacobi
                field of geodesic separation.
\bibitem{nota3} This is a consequence of the possibility of approximating
                locally the metric of a manifold by $g_{ik}\simeq\delta_{ik}-
\frac{1}{6}R_{ikjl}u^iu^k$ obtained by using normal coordinates and 
displacements $u^i$ (see e.g. Do Carmo, Ref. \protect\cite{DoCarmo}).
\end{references}
\end{document}